\documentclass{aa}   
\usepackage[varg]{txfonts}
\pdfoutput=1 
 
\usepackage{amsmath}
\usepackage{amsfonts}
\usepackage{amssymb}
\usepackage{graphicx}
\usepackage{fourier}

\usepackage{natbib}
\bibpunct{(}{)}{;}{a}{}{,} 

\usepackage{booktabs} 

\usepackage{makecell}
\usepackage{txfonts}
\usepackage{graphicx,natbib,url,twoopt}
\usepackage{lipsum}
\usepackage{multicol}
\usepackage{float}

\usepackage{threeparttable}
\usepackage{adjustbox}

\usepackage{caption}
\usepackage{subcaption}

\def\m2s2{\hbox{\,m$^{2}$\,s$^{-2}$}} 

\def\mearth{\hbox{$\mathrm{m}_{\oplus}$}}     

\def\mjup{\hbox{$\mathrm{m}_{\rm Jup}$}}

\begin{document}
\title{A general stability-driven approach for the refinement of multi-planet systems\thanks{The new HARPS and CORALIE data of HD\,45364 presented in Appendix \ref{Appendix:HD45364_DataAnalysis} are only available in electronic form at the CDS via anonymous ftp to \protect\url{cdsarc.u-strasbg.fr} (\protect\url{130.79.128.5}) or via \protect\url{http://cdsweb.u-strasbg.fr/cgi-bin/qcat?J/A+A/ }.} }

\author{M. Stalport
\and J.-B. Delisle
\and S. Udry
\and E. C. Matthews
\and V. Bourrier
\and A. Leleu
} 

\institute{D\'epartement d'Astronomie, Universit\'e de Gen\`eve, Chemin Pegasi 51b, CH-1290 Versoix, Suisse \label{i:geneva}}
 
\date{Received ?? / Accepted ??} 
 
\abstract
{Over the past years, the amount of detected multi-planet systems significantly grew, an important sub-class of which being the compact configurations. A precise knowledge of them is crucial to understand the conditions with which planetary systems form and evolve. However, observations often leave these systems with large uncertainties, notably on the orbital eccentricities. This is especially prominent for systems with low-mass planets detected with Radial Velocities (RV), the amount of which is more and more important in the exoplanet population. It is becoming a common approach to refine these parameters with the help of orbital stability arguments.}
{Such dynamical techniques can be computationally expensive. In this work we use an alternative procedure faster by orders of magnitude than classical N-body integration approaches, and with the potential to narrow down the parameters' uncertainties.}
{We couple a reliable exploration of the parameter space with the precision of the Numerical Analysis of Fundamental Frequencies \citep[NAFF,][]{Laskar1990} fast chaos indicator. We also propose a general procedure to calibrate the NAFF indicator on any multi-planet system without additional computational cost. 
This calibration strategy is illustrated on HD\,45364, in addition to yet-unpublished measurements obtained with the HARPS and CORALIE high-resolution spectrographs. We validate the calibration approach by a comparison with long integrations performed on HD\,202696. 
We test the performances of this stability-driven approach on two systems with different architectures. First we study HD\,37124, a 3-planet system composed of planets in the Jovian regime. Then, we analyse the stability constraints on HD\,215152, a compact system of four low-mass planets.}
{We revise the planetary parameters in HD\,45364, HD\,202696, HD\,37124 and HD\,215152, and provide a comprehensive view of the dynamical state these systems are in.}
{We demonstrate the potential of the NAFF stability-driven approach to refine the orbital parameters and planetary masses. We stress the importance of undertaking systematic global dynamical analyses on every new multi-planet system discovered.}


\maketitle
\section{Introduction} \label{Intro} 
After the first detection of an exoplanet orbiting a main-sequence star by \citet{Mayor1995}, it did not take long to discover the first multi-planet system \citep[Upsilon Andromedae,][]{Butler1999}. Since then, the number of discovered multi-planet systems has significantly increased\footnote{823 multi-planet systems were known as of May 6, 2022 (from \url{http://exoplanet.eu/catalog/}).}. Dedicated exoplanet missions such as Kepler and TESS or results from high-resolution spectrographs such as HARPS boosted this amount, and illustrated the large diversity of architectures that are observed. In the solar neighbourhood, the planetary orbits in multi-planet systems harbour very different shapes and levels of compactness. This observed diversity provides important constraints on the planet formation and orbital evolution models. Indeed, the current planetary orbits act as tracers of their birth place and past orbital evolution under the influence of planet-disk and/or planet-planet gravitational interactions. Furthermore, the direct comparison of the orbital configurations and properties of the planets within the same system provides indications in regards of the formation processes \citep[e.g.][]{MacDonald2016, Hadden2020, Leleu2021, Heising2021}. Multi-planet systems therefore represent precious case studies to help understanding the formation of planetary systems as a whole. Among the most interesting multi-planet systems, compact architectures are particularly intriguing. With no equivalent in the solar system, they are subject to strong dynamical constraints. Those systems thus represent ideal candidates to test the role of planet-planet gravitational interactions in the formation and evolution of planetary systems \citep[e.g.][]{Hadden2020}. However, as those systems are often composed of small planets, they are also observationally challenging. As a result, such observations are likely to provide poor constraints on the orbital and planetary properties in these systems.  

The study of dynamics, and in particular orbital stability, has been a well-established technique to refine the orbital parameters of multi-planet systems. A common procedure is to build chaos maps - also sometimes named stability maps, which explore the chaotic behaviour of 2-dimensional sections of the parameter space \citep[e.g.][]{Gozdziewski2001, Correia2009, Couetdic2010, Lovis2011, Robertson2012, Satyal2014, Wittenmyer2014, Delisle2018, Hadden2020, Leleu2021}. They provide visual insights into the structure of the parameter space and precise the dynamical state of the studied system. However, such a procedure has limitations, because it gives a partial view of the parameter space. Indeed, chaos maps explore the influence of two parameters only, while the whole parameter space of the multi-planet system can be of large dimension. In particular, they do not account for potential correlations between parameters. As such, chaos maps represent a restricted exploration of the parameter space, and one has to be cautious with their interpretation. 
In order to overcome this limitation, dynamical methods that include the stability information during the fitting procedure have been shown very efficient. Notably, 
a technique coupling a genetic algorithm for the parameter space's exploration with the MEGNO fast chaos indicator to discriminate the strongly chaotic solutions was proposed \citep{Gozdziewski2003,Gozdziewski2008}. This approach, named GAMP (for Genetic Algorithm with MEGNO Penalty), allows for a global exploration of the parameter space and an efficient estimation of orbital stability. The latter is based on heuristic arguments, coupling a level of chaos to a level of stability. This technique proved able to constrain the system when only a few Radial Velocity (RV) measurements are available.

A particularly well-indicated and well-established method of parameter space's exploration is the Markov Chain Monte Carlo (MCMC) algorithm. It consists in a random walk inside the parameter space, yet influenced by some prior information and by the constraints that the observational data add. The resulting path of the walker defines a global posterior distribution of the same dimension than the parameter space. The projection of this global posterior on each parameter serves as the baseline for the parameter's estimation: we often use the median of the distribution together with its 68.27$\%$ confidence interval. This technique is very widely used among the community to estimate model parameters from observational data. 
As a result of its popularity and efficiency, there is an interest to couple the MCMC exploration tool with the orbital stability constraint. 
This general approach is not new. To our knowledge, its first appearance dates back in 2006, when \citet{Ford2006} proposed to use N-body numerical simulations to remove from the posterior distribution all the solutions that turned unstable over the course of the integrations. This importance sampling based on orbital stability estimation is interpreted as another constraint on the system, in addition to the observations. This strategy was applied in \citet{Veras2010}, which performed an exhaustive dynamical analysis on a set of five multi-planet systems. For each of them, they modelled the radial velocity measurements and explored the parameter spaces of those models with MCMC. They carried on N-body numerical simulations over 1 Myr on a sampled posterior for each system. Besides close encounters, their stability criterion was set from the variations in the planets' semi-major axes, which shouldn't exceed a defined threshold. 
Similar approaches were employed later on, with the same aim to provide a rigorous revision of the orbital parameters \citep[e.g.][]{Joiner2014, Nelson2016, Trifonov2019, Quinn2019}. 

In this paper, we develop a technique coupling the efficiency of the MCMC to explore the parameter space with a significantly more rapid stability estimation to perform rejection sampling. 
This estimation is based on a heuristic approach associating the physical stability (i.e. close encounter or escape) to a measurement of chaos. In particular, it makes use of the Numerical Analysis of Fundamental Frequencies \citep[NAFF, ][]{Laskar1990, Laskar1993} fast chaos indicator. The latter quantifies the chaos of a planetary system by measuring the drift in the average mean motions of the planetary orbits over time. While in the secular theory, non-chaotic orbits do not display any drift in the average mean motions, this is not true anymore in chaotic trajectories. Furthermore, the larger are the drifts, the more chaotic are the orbits and hence, the more unstable is the planetary system.  Nevertheless, a major drawback of this indicator in the context of this work is the lack of general calibration for the drift in the average mean motions. This issue is tackled here. 
This procedure linking bayesian techniques to explore the parameter space with the NAFF fast chaos indicator for the estimation of orbital stability was already used in previous studies with the aim of refining the orbital elements and masses in multi-planet systems \citep{Nielsen2020, Hara2020}, without however the use of a universal NAFF calibration procedure. In this paper, we present the method and demonstrate its performances.  

We first stress the link between orbital instability and chaos and introduce the NAFF indicator in Sect. \ref{Section:StabIndic}.  We propose a strategy to calibrate the drift in average mean motion that the NAFF measures (Sect. \ref{Sect:NAFFcalib}). The approach is illustrated with HD\,45364, a system of two massive planets in the 3:2 MMR. We validate the calibration procedure on HD\,202696, by comparing the results of the fast chaos estimation approach with brute-force long integrations. Then, we apply the stability-driven refinement technique on HD\,37124 in Sect. \ref{Sect:HD37124}. This three-planet system has already been studied in light of its orbital dynamics, but so far no exhaustive revision of the parameters has been led. Our stability-driven refinement technique is also tested on a compact system of low-mass planets: HD\,215152 \citep{Delisle2018}. This system is one of the few compact configurations found with RV measurements only. The observations do not constrain well the orbital eccentricities, which leaves room for the refinement technique to play a significant role in updating the system parameters rigorously. This revision is presented in Sect. \ref{Sect:HD215152}. Finally, Sect. \ref{Sect:Conclusions} discusses our results.

\section{Method}
\label{Sect:Method}

\subsection{Strategy of the approach} 
\label{SubSect:Strategy} 
There are three general strategies with which one can add the stability information onto the results of the MCMC exploration. 
First, the stability information could be added as a prior information for the MCMC, based on either analytical arguments or trends in the observed multi-planet systems. However, so far the numerical integrations bring stronger constraints on the systems than any of these arguments. 

Secondly, it could also be included during the MCMC exploration, as a second likelihood information together with the observational data. Such a procedure demands some caution though. Indeed, the amount of steps a walker typically performs in a MCMC is on the order of the million or more, and thus the orbital stability of as many system's configurations should be estimated. The use of analytic stability criteria is well-indicated for that, since they require very small computing times. But they are less reliable than the integration techniques, and are furthermore restricted to specific orbital architectures, which is not convenient for a general purpose. Recently, a fast stability estimation based on short integration times together with machine learning patterns recognition has shown promising performances. This tool, named SPOCK \citep{Tamayo2020SPOCK}, is however dependent on the training set for the machine learning algorithm. With an increasing diversity of systems' architectures used during the training phase, and given an efficient computing cluster at disposal, this tool constitutes a very competitive alternative to the analytical criteria. 

Thirdly, the orbital stability constraint could be added after the MCMC exploration. The orbital stability would be estimated on each system's configuration of the posterior sample - or a sub-sample of the latter. A disfavouring weight on the unstable configurations would re-shape the posteriors. 
Such a strategy is analogous to importance sampling. It is effective if enough stable samples remain in the distribution at the end of the process. As a result, the less constrained a system is, the larger the distribution should be in order to correctly explore the stable areas of the parameter space. 
Compared to the second strategy though, that one necessitates the estimation of stability on much less system's configurations, and can thus be coupled with classical integration techniques. Furthermore, it allows stressing the impact that the dynamical constraints add on the system, by directly comparing the posteriors before and after the addition of the stability computation. Therefore, we opt for this third strategy in this work. 
Our stability-driven approach thus combines the posterior of global model parameters exploration with a decisional process for stability.

\subsection{Choice of the stability criterion}
\label{Section:StabIndic}
In many studies using stability arguments a posteriori of the MCMC, the stability proxy is simply the survival of the system's configuration over the entire timespan of the integration \citep[e.g.][]{Trifonov2019, Quinn2019, Cloutier2019}. Such a criterion is very reliable against the unstable behaviour of the configuration. However, it does not give any information about the orbital stability after the integration ends. As such, somewhat large integration times are needed for the stability decision to make sense, i.e. over a timespan not too small compared to the age of the planetary system under study. Thus, it is computationally expensive. There exists though alternative options of stability criteria. A convenient choice is to use chaos indicators, which proved efficient in spotting orbital instabilities ahead of their appearance. 

Chaos is a feature of the motion by which a dynamical system will harbour uneven, disordered evolution and with large variations in the trajectories. A consequence of these features is a hyper-sensitivity regarding the initial conditions from which the system is evolving. 
Many different chaos indicators were built, with varying efficiencies - e.g. the well-established Lyapunov Characteristic Exponents \citep[][for a review]{Benettin1980}, the Numerical Analysis of Fundamental Frequencies NAFF \citep{Laskar1990, Laskar1993}, the Spectral Number SN \citep{Michtchenko1995, Michtchenko2002}, the Conditional Entropy of Neary Orbits \citep{Nunez1996}, the Fast Lyapunov Indicator FLI \citep{Froeschle1997}, the Mean Exponential Growth of Nearby Orbits MEGNO \citep{Cincotta2003}, the 0-1 test \citep{Gottwald2004}, the Shannon entropy \citep{Giordano2018,Cincotta2021}. 
Their use is not restricted to planetary dynamics, but has had also applications to e.g. the stability of stellar orbits in galactic potentials \citep{Udry1988}. 
All of these indicators find their basic principle in the link between chaoticity and instability \citep[e.g.][]{Chambers1996, Murray1997, Obertas2017, Rice2018, Hussain2020}. This observed link has limitations though, as the information of one quantity does not always rely on the other, notably when a dynamical system harbours what is called bounded chaos. In those systems, the topology of the parameter space is such that the chaotic area is spatially restricted, and the system can only diffuse between impervious boundaries. As a result, the dynamical system may harbour strong chaos but stable behaviour, because of the very limited impact on the orbital elements' variations. 

Besides these particular cases and although no explicit general relationship has been validated between chaoticity and instability time, it is nowadays generally accepted and empirically proved that the more chaotic is a system's configuration, the more unstable it will be, i.e. the shorter will be its lifetime. Extensive numerical simulations of the chaotic behaviour of planetary systems furthermore showed that the stronger the chaos, the smaller the uncertainty on the survival time \citep{Hussain2020}. This result implies that, due to intrinsic uncertainties coming from the chaotic nature of motion, it becomes highly doubtful to associate a moderately chaotic system with a certain survival time. Instead, in this work we will focus on strong chaos, from which a following instability can be inferred. Our approach aims at reaching similar results than the "brute force" long integrations technique, with shorter by orders of magnitude integration times.  

A sub-class of chaos indicators is particularly well-suited, which is the fast chaos indicators. They converge rapidly to a stability answer, and thus do not necessitate long integration times. Among them, the Numerical Analysis of Fundamental Frequencies \citep[NAFF, ][]{Laskar1990, Laskar1993} has been widely used among the community. 
While most of the other mentioned chaos indicators estimate chaos from the differential evolution of two initially nearly-identical system's configurations - exploiting the hyper-sensitivity to initial conditions of chaotic orbits -  the NAFF finds its basics on a technique named frequency analysis. The later was first presented in \citet{Laskar1988} for the computation of the secular evolution of the solar system via a semi-analytical approach. It is a tool that estimates with high precision the main frequencies in a dynamical system, with the hypothesis that the orbits are quasi-periodic. The NAFF indicator uses this tool in order to test the quasi-periodic hypothesis of the orbital motion. 

Generically in any coordinate system, each coordinate $f$ of a quasi-periodic trajectory can be decomposed as: 
$$ f(t) ~ = ~  \sum_{k=1}^{\infty} A_k ~ e^{i \nu_k t} $$
which is an infinite sum of harmonics, with frequency terms $\nu_k$ and amplitudes $A_k$, $t$ being the time and $i$ the imaginary unit. The aim of frequency analysis is to numerically approximate $f$ by searching for the $N$ highest amplitude terms of the series, with the help of numerical integration. Thus, we build a function $f'$ that numerically approaches $f$, in order to get 
$$ f'(t) ~ = ~  \sum_{k=1}^{N} A'_k ~ e^{i \nu'_k t}. $$ 
The procedure to get the approximative function $f'$ for the motion of a planet starts by integrating the whole system over a time $T$. The planetary motion over that timespan is then split in its constituent frequencies with the help of a refined Fourier analysis. The frequency with the largest amplitude is removed, and a new spectrum is computed. This process is repeated $N$ times in order to provide the $N$ highest amplitude terms of the decomposition. This is an iterative process. 

Frequency analysis is used to estimate the frequency associated with the mean longitude, which is the mean motion $n=\frac{2\pi}{P}$. That frequency appears always as the first term of the above decomposition because its amplitude is the highest. 
We thus use the frequency analysis with $N$=1, and there is no confusion about the physical origin of that term. Furthermore, in the quasi-periodic hypothesis of motion, the mean motion does not harbour any secular variations or drifts. It is constant. The NAFF stability indicator precisely tests this hypothesis, by searching for potential drifts in the value of the mean motion. Because the mean longitude is a quickly varying angle, this coordinate is sensitive to chaos on short timescales, which allows us to build a fast chaos indicator. The procedure employed by NAFF to do so is explained in section 9 of \citet{Laskar1990} and applied to exoplanet systems in e.g. \citet{Correia2005}. The idea is to split the total integration time in two halves and apply the frequency analysis technique on each half. This process will provide us twice with a very precise estimation of the fundamental frequency associated with the mean motion of the planet on its orbit. The comparison between these two frequencies is at the basis of the chaos estimation. The smaller this difference, the closer to real quasi-periodicity is the orbit and thus the less chaotic is the trajectory. However, if the difference is larger, the orbit displays stronger chaos and hence increases the probability of instability. 

In this work, we will compute the drift in mean motion of every planetary orbit. In order to compare that drift between planetary orbits of different scales, we normalise it with the initial mean motion of the planet under study. As such, we define our NAFF stability indicator as the following:  
\begin{equation} \label{eq:NAFF}
\mathrm{NAFF} ~ = ~ \max_{j} ~ \left[ log_{10} \dfrac{\mid n_{j,2} - n_{j,1} \mid}{n_{j,0}} \right] 
\end{equation}
where $n_{j,1}$ and $n_{j,2}$ are the estimated mean motions over the first and second half of the integration respectively, for every planet $j$ in the system. $n_{j,0}$ is the initial Keplerian mean motion of planet $j$. We retain the maximal drift value among all the planets as the proxy for the chaos in the system. The convergence of this chaos indicator is dictated by the precision in the estimation of the mean motion frequencies. The latter directly depends on the integration time, and usually reaches reliable results after $\sim10^4$ revolutions of the outermost planet.  

Let us note that we will use the NAFF indicator as a binary decision maker about orbital stability. Therefore, we set up a NAFF threshold above which the averaged mean motion variations are considered too large for the system to be stable on long time scales. In Sect. \ref{Sect:NAFFcalib} we propose a technique to find an appropriate NAFF threshold according to the planetary system under study. This approach does not necessitate further numerical simulations. Another option for the use of NAFF would have been to assign a weight to each realisation of the MCMC posterior inversely proportional to the average mean motion drift. This weight would be added to the function informing about the quality of the fit, e.g. the $\chi^2$. 
However, we remind that we use NAFF to spot the strongly chaotic system's configurations, for which the drifts in mean motion are larger. A smooth weight function would thus have its domain in the large NAFF values only, while all the small NAFF values would be assigned a weight of 1. We empirically noticed that such a function is well-approximated with a strict cut. 

\begin{figure}
    \centering
\includegraphics[width=\columnwidth]{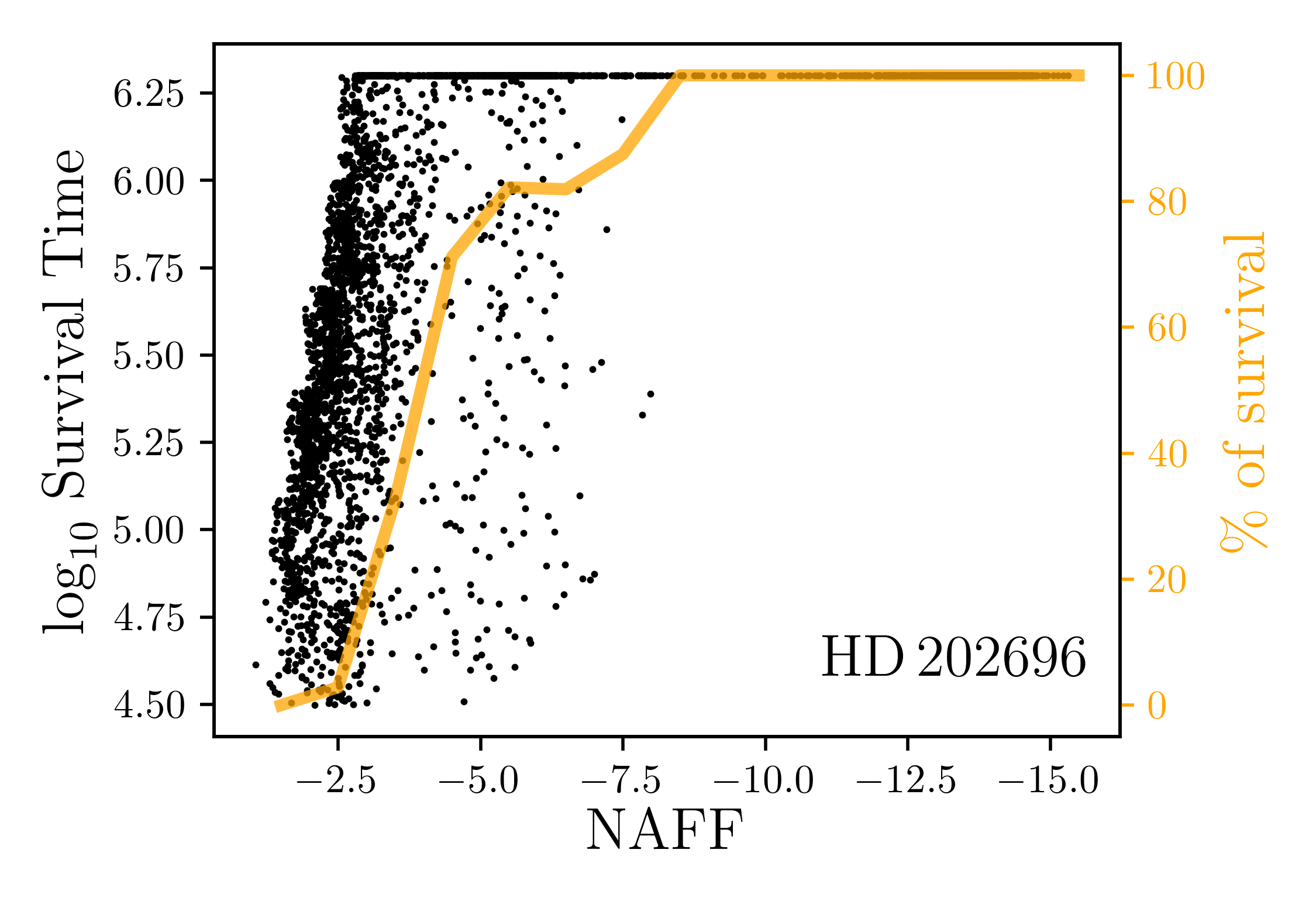}
    \caption{HD\,202696. \textit{Black}: Scatter plot of the NAFF chaos indicator with respect to the actual survival time of the system's configurations. \textit{Orange}: Proportion of stable systems (i.e. that survived the entire 2 Myr of integration) for consecutive ranges of NAFF.}
    \label{fig:HD202696_NAFF-ST}
\end{figure}

In Fig. \ref{fig:HD202696_NAFF-ST}, we illustrate this by comparing on the MCMC posterior the results of the NAFF chaos indicator with the actual survival time of the system in the case of the two-planet system HD\,202696. The two massive planets reported in \citet{Trifonov2019} - minimum masses of 1.99 and 1.92\,\mjup for planets b and c, respectively - orbit the star with a period ratio of $\sim$\,1.81 ($P_b\sim$528.2 days, $P_c\sim$958 days). We fitted the 42 HIRES radial velocity data publicly available for this star (M=1.91M$_{\odot}$). This was done using the Data $\&$ Analysis Center for Exoplanets (DACE) platform\footnote{The DACE platform is a facility of the National Center of Competence in Research PlanetS based at the University of Geneva dedicated to extrasolar planets data visualisation, exchange and analysis. It is available at \url{https://dace.unige.ch}.}: first via an iterative signal search in the periodogram of the RV timeseries, then modelling the significant periodicities with keplerians, and finally by computing the periodogram of the residuals to restart again until no significant peak is left. At the end of this process, we used a MCMC to efficiently explore the parameter space of the model. The implementation of this MCMC is described in \citet{Diaz2014}, and is based on a Metropolis-Hastings algorithm. We removed the first 25$\%$ of the iterations as a burning phase, and sampled the posterior according to the correlation length of the chain. We then integrated every configuration of the resulting MCMC posterior distribution over 2 Myr, i.e. twice the integration time used in the reference study by \citet{Trifonov2019}. The numerical simulations were performed in the centre-of-mass inertial frame, and using the 15th order adaptive time-step integrator \texttt{IAS15} \citep{Rein2015}. The latter is based on a Gauss-Radau quadrature and is implemented in the \texttt{REBOUND} python package\footnote{\texttt{REBOUND} is a package to simulate with N-body the motion of particles under gravitational interactions. It harbours a wide variety of numerical integrators, and can be found at \url{https://rebound.readthedocs.io/en/latest/}.} \citep{Rein2012}. We also computed the NAFF chaos indicator on the first 31\,250 years only, i.e. a 64th of the total integration time\footnote{NAFF convergence checks on this system were performed on 64 successive integration times, starting at 31\,250 years. The convergence was already good with this first time span, which explains the choice of 31\,250 years for the NAFF computation.}. 
Out of the 7\,813 system's configurations making the MCMC posterior, 4\,688 reached the first 31\,250 years of integration and hence got a NAFF estimation 
- the others got unstable (close encounter or escape) before. To proceed at the NAFF chaos computation, we recorded the planets' mean longitudes at regular time steps for a total of 20\,000 output times. From these timeseries, we can decompose every planetary motion into its constituent frequencies and apply the frequency analysis technique. This serves at computing the average mean motion of every planet over each half of the integration, and derive the NAFF chaos indicator (eq. \ref{eq:NAFF}). 

The numerical estimation of the frequencies obtained with the NAFF algorithm is affected by the output time step $\delta t$ as well as the total duration $T$ of the integration. The precision on the frequencies is mainly limited by the total duration of the integration \citep{laskar2003}. On the other hand, the time step $\delta t$ has a weak impact on the precision of the frequencies but introduces aliases issues. Indeed, one cannot distinguish between a frequency $n$ and a frequency $n + k n_\mathrm{Nyquist}$, where $k$ is an integer and $n_\mathrm{Nyquist} = \pi/\delta t$ is the Nyquist's frequency. With our choice of sampling in the HD\,202696 study, we actually end-up in a case where the time step is longer than the period of the outer planet, which means that the frequency we obtain with the NAFF algorithm is actually an alias of the mean motion. However, we are mainly interested in the difference in the mean motion between the two halves of the integration, which is given by:
$n_2 - n_1 = f_2 - f_1 + (k_2-k1) n_\mathrm{Nyquist}$, where $f_1$ and $f_2$ are the frequencies obtained through the NAFF algorithm. We then assume that the mean motion did not dramatically change and we select the value of $k = k_2 - k_1$ that minimises $n_2-n_1$. This procedure works as long as the mean motion drift is small compared to Nyquist frequency. 

The scatter graph of Fig. \ref{fig:HD202696_NAFF-ST} represents the 4\,688 configurations that benefit from a NAFF estimation after 31\,250 yr. Each of these is a dot in the space of NAFF versus the survival time of the system's configuration, in $\log_{10}$ scale. Naturally, the maximum survival time that we can estimate is the total integration time, i.e. 2 Myr. 
As can be seen in the figure, a proportion of configurations reached that limit, and build up the horizontal series of points on the top. 
Despite some spread inherent to the chaotic nature of the systems, we observe a clear correlation in the strong chaos regime. Starting from NAFF $\sim$ -2.5, the first configurations that survived the full integration appear, and this proportion increases with decreasing NAFF. In the figure we also show in orange, for consecutive NAFF ranges, the proportion of systems that survived the entire 2 Myr integration. A value of 100$\%$ means that all the system's configurations in a specific NAFF range are stable - in the sense that they survived the entire 2 Myr of integration (no escape, no close encounter), while a value of 0$\%$ indicates that none reached the end. This curve would represent our weight function. 
However, given its sharp increase in the more chaotic domain, we notice that a strict cut above a certain NAFF threshold is close to the weight function. Hence, we decide to use NAFF as a binary stability indicator, which considerably simplifies its use.

\section{NAFF calibration}
\label{Sect:NAFFcalib}
The drift in the mean motions provides a relative measurement of chaos. Indeed, despite the normalisation performed in eq. \eqref{eq:NAFF}, it is not granted that two different systems with the same maximal drift in the mean motions of the planetary orbits will have the same survival time. In other words, the NAFF indicator does not provide any information regarding the absolute chaos - the same amount of NAFF may correspond to different levels of orbital stability among various system's architectures. However in this work, we will make use of NAFF as a binary stability decision maker. In order to set up a consistent stability threshold, a calibration of NAFF for every studied system becomes necessary. 

To our knowledge, very few studies attempted to calibrate the NAFF chaos indicator. At first, linear diffusion laws were hypothesised to extrapolate the evolution of the mean motion drifts \citep[e.g.][]{Robutel2001}. Such an extrapolation could be used to predict the time after which the mean motions had varied by more than 100$\%$, and use this time estimation in a similar fashion as the Lyapunov time. This process however is approximative, as it relies on the hypothesis of linear growth in the mean motion drift. A more rigorous process of NAFF calibration was proposed by \citet{Couetdic2010}. Their strategy consists in defining a grid of systems, similar to a grid used for chaos maps, and compute the NAFF chaos indicator (without normalisation in their case) of each system for two integration times $T_1$ and $T_2$ ($T_1$>$T_2$). If a system was quasi-periodic, its NAFF estimation should get smaller over a longer integration timespan $T_1$, because of the better numerical precision on the estimation of the fundamental frequencies. For chaotic systems though, the chaotic diffusion would dominate and we would find that NAFF$_1$>NAFF$_2$. To separate these two regimes, the authors select the NAFF threshold as the approximate value NAFF$_1$ at which 99$\%$ of the systems harbour NAFF$_1$<NAFF$_2$. This percentage was selected in order to minimise the amount of false positives and false negatives, by comparison with long numerical integrations. This 99$\%$ value may not be convenient for significantly different system's architectures. To verify it, one would need to perform additional long numerical integrations for any new system under study. 
In this section, we propose a general NAFF calibration technique which has the advantage that no additional numerical integration is required.

In the analysis of several multi-planet systems, it has been observed that adding new observations shifts the position of the derived solution of the system in the chaos maps towards zones of more regularity - e.g.  GJ\,876 \citep{Correia2010}, HD\,45364 \citep{Correia2009}, HD\,202206 \citep{Couetdic2010}, HD\,60532 \citep{Laskar2009}. This is in agreement with the work hypothesis that observed planetary systems are stable. That hypothesis is justified by the rapidity with which orbital instabilities develop \citep[e.g.][]{Petit2020}. Hence, unstable configurations would most probably have already been destroyed over the typical $\sim$Gyr timescale of the system's lifetime. Linking this with the previous observation, it means that every observed system should not be intrinsically strongly chaotic, and that the addition of observational constraints moves the systems deeper inside zones of regularity. When we model a system from observational data, the ensemble of configurations that compose the posterior distribution often harbours two types of behaviours. Some show weak chaos, others strong chaos. As a result of what was said above, the more constraining are the data - either because of higher precision in the individual measurements, or because of an increasing amount of data points - the larger would be the weak chaos sub-population compared to the strong chaos one. 

We illustrate this point and exploit this observation to calibrate the NAFF chaos indicator. 

\subsection{Illustration with HD\,45364}
\begin{figure}
    \centering
\includegraphics[width=\columnwidth]{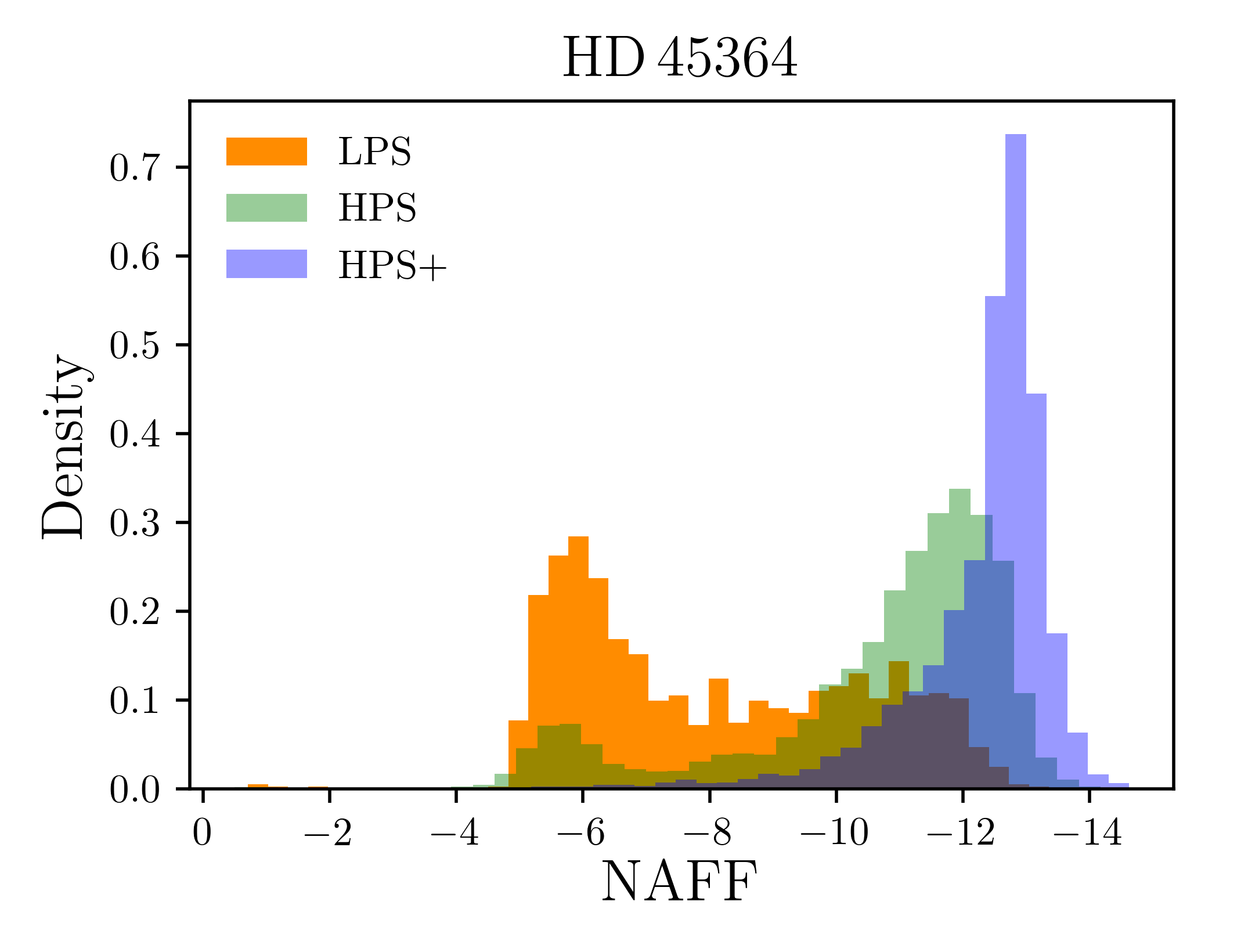}
    \caption{HD\,45364: Comparison between the NAFF distributions of Low and High Precision Sets of configurations (LPS, HPS and  HPS+, respectively). The HPS+ dataset adds yet-unpublished RV data from both CORALIE and HARPS spectrographs.}
    \label{fig:DistribNAFF_HD45364}
\end{figure}

\begin{figure*}
    \centering
\includegraphics[width=\textwidth]{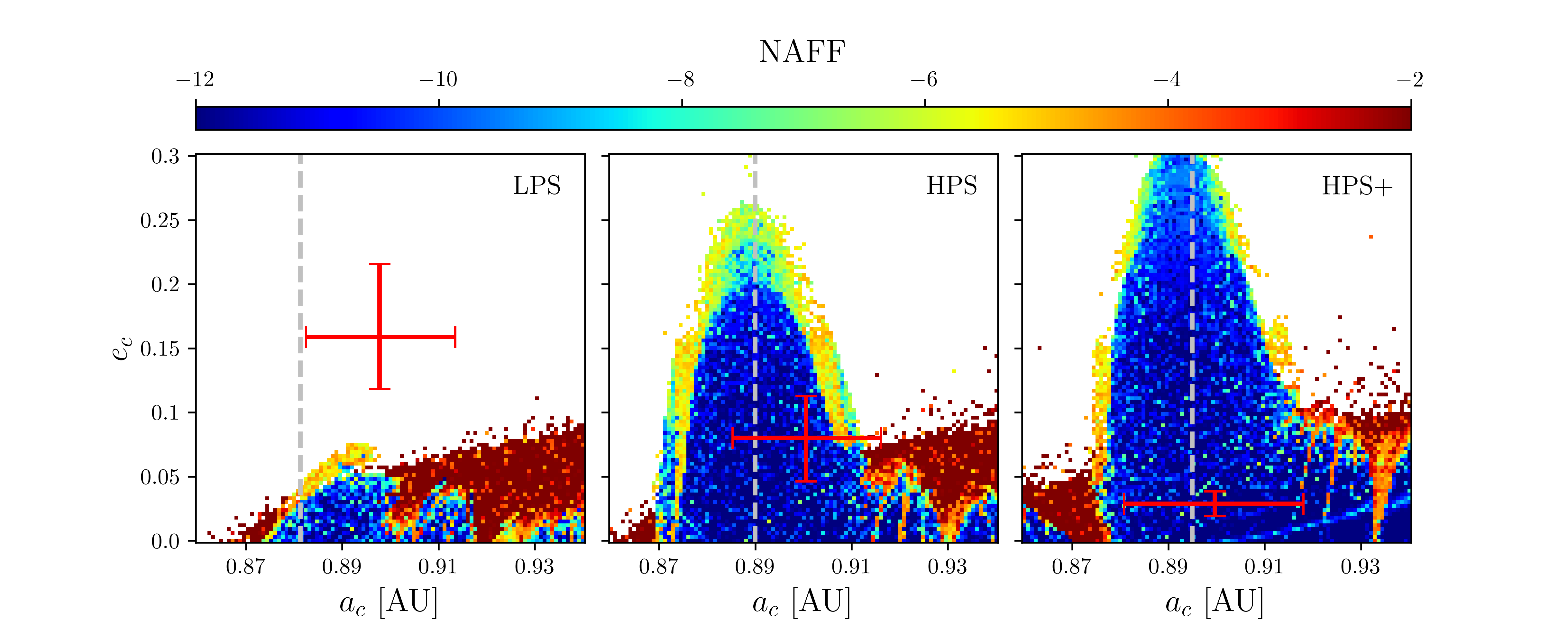}
    \caption{HD\,45364: Chaos maps around the 3:2 MMR in the sub-space ($a_c$,$e_c$) for each set of RV measurements. The red crosses represent the 1$\sigma$ dispersion around the median value of the MCMC posterior distribution for both $a_c$ and $e_c$. The vertical dashed lines depict the location of the periods' commensurability ($P_c/P_b$=3/2), which changes from a map to another with different estimations of $P_b$. The colour code depicts the NAFF chaos level, while the squares are coloured in white if the corresponding system's configuration did not reach the end of the simulation (either because of a close encounter or an escape). From left to right: LPS, HPS and HPS+ datasets.}
    \label{fig:StMaps_HD45364}
\end{figure*}

HD\,45364 is composed of a pair of giant planets with orbital periods of $P_b$ = 226.9 days and $P_c$ = 342.9 days, and minimum masses of 0.19 and 0.66 \mjup. The discovery was reported by \citet{Correia2009} using 58 HARPS spectra. These authors showed the system to lie in the 3:2 MMR, surrounded by a sea of strong chaos (cf. Fig. 4 in their article). Later on, additional studies were led in order to infer the formation and migration history in a proto-planetary disk based on the resonant configuration \citep{Rein2010,Correa-Otto2013,Delisle2015,Hadden2020}. In this work, we analyse three datasets of different sizes, and compare them in terms of the chaoticity of their solutions. First, we define the Low Precision Set (LPS) that contains only 33 RV measurements taken from the published RVs in \citet{Correia2009} and removing the last part of the timeseries. Secondly, we define the High Precision Set (HPS) which consists of all the 58 measurements used in the discovery paper. Finally, we introduce a third dataset composed of the 58 published RVs together with additional yet-unpublished RV measurements coming from both the CORALIE and HARPS spectrographs and taken between Jan 2001 and Feb 2021. Altogether, we add 57 measurements from HARPS and 68 from CORALIE, for a total of 183 RV measurements. The presentation of this new timeseries and its analysis with Lomb-Scargle periodograms is provided in Annex \ref{Appendix:HD45364_DataAnalysis}. This defines the High Precision Set + (HPS+). 

For each of these three datasets we nightly bin the RV measurements, fit the RV timeseries with a two-keplerian model and explore the parameter space of the latter with an MCMC algorithm - all of this using the DACE platform. Thus for the LPS, HPS and HPS+ we derive global posterior distributions for our model parameters, from which we construct sub-samples of 6\,250 solutions. We then numerically integrate each solution over 50 kyr with the \texttt{IAS15} integrator. The results of these integrations, when they reached the end (neither close-encounter nor escape), are used to compute the NAFF chaos indicator of the configurations. To do so, we output from each integration an evenly sampled series of 20\,000 mean longitude values for each planetary orbit, on which we apply the frequency analysis technique. 

In Fig. \ref{fig:DistribNAFF_HD45364} we present the NAFF histograms for these three distributions: the LPS in orange, the HPS in green and the HPS+ in blue. The histograms corresponding to the LPS and HPS datasets clearly show two different dynamical populations: the population of stronger chaos on the left, and the population of smaller mean motion drift and hence higher regularity on the right. The set of configurations obtained with less data (LPS) presents an asymmetry in favour of the more chaotic population, while the NAFF histogram of the HPS displays the opposite division. It is reasonable to expect that the addition of even more data would ultimately result in the total absence of the more chaotic configurations, and leave the MCMC posterior with only weakly chaotic dynamical behaviour. That is what is observed when we add the yet-unpublished data: the MCMC posterior from the HPS+ set only contains system's configurations in the regular weak chaos regime. Hence the new observations further constrain the system to lie in the stable 3:2 MMR island. As an illustrative counterpart, we also computed chaos maps around the 3:2 MMR in the sub-space ($a_c$,$e_c$) for each of the RV datasets. They are presented in Fig. \ref{fig:StMaps_HD45364}. For each map, 101$\times$101 system's configurations are generated and their future evolution are numerically computed over 50 kyr using the same set-up. The NAFF indicator is estimated on each configuration that survived the entire integration, via the same strategy as explained above. Essentially two phenomena are observed in this suite of chaos maps. The estimation of the orbital eccentricity $e_c$ is converging towards small values, while the semi-major axis $a_c$ gets closer and closer to the 3:2 period commensurability. The crosses inform about the area of the parameter space sampled by the MCMC. The addition of more data constrains this exploration to the stable resonant island. This explains the behaviour observed in Fig. \ref{fig:DistribNAFF_HD45364}. It is important to remind that while the other orbital parameters are fixed initially in each map, they vary from one map to another as they are updated with the addition of RV measurements. They also converge towards the stability island, explaining why it appears more and more clearly with the addition of RV data. 

\begin{figure*}[t!]
    \centering
\includegraphics[scale=0.7]{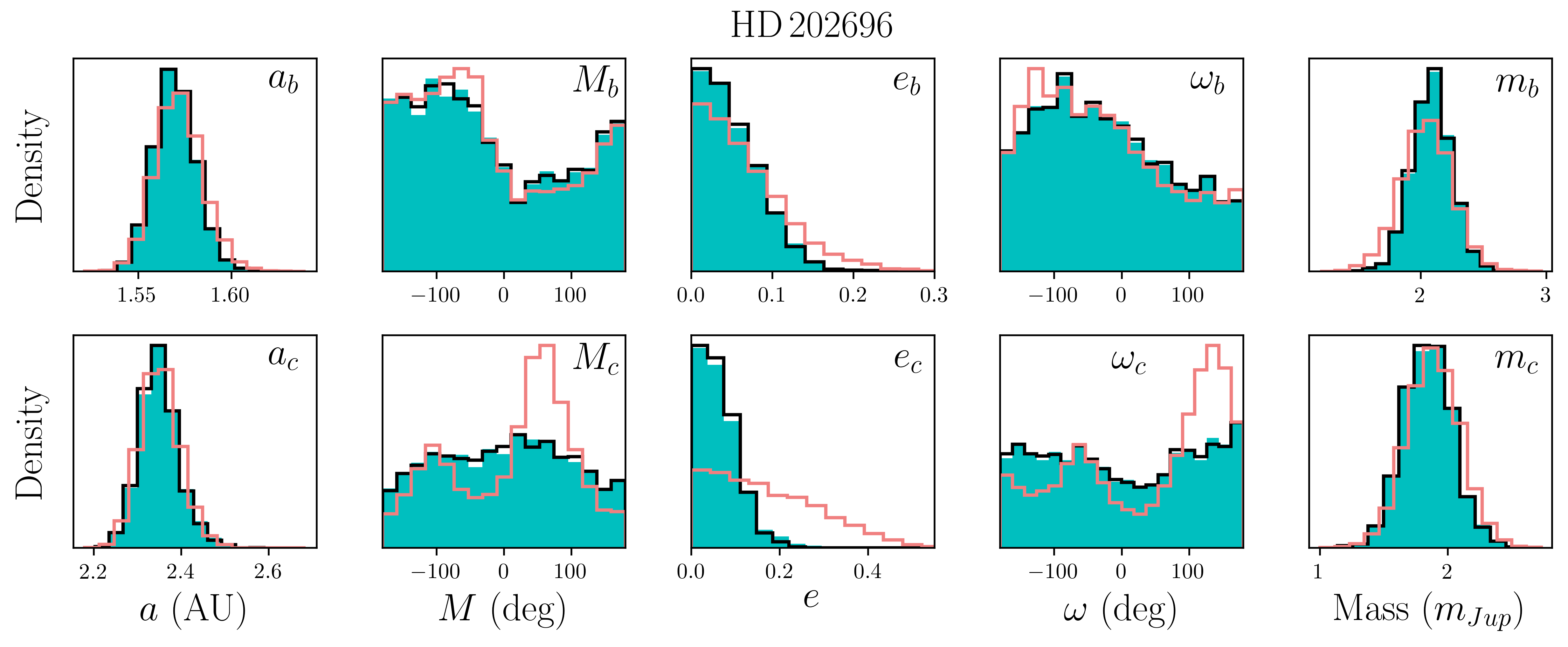}
    \caption{HD\,202696: Comparison between the posterior distributions generated from three different processes, for each planetary dynamical parameter. In red: the posterior distributions as obtained from the observational data only. In black: the sub-set of stable configurations as defined by the "brute force" long integrations. In blue: the NAFF-stable distributions. Every system's configuration is supposed co-planar, leading to undefined longitudes of the ascending nodes $\Omega$.}
    \label{fig:HD202696_DistribCompa}
\end{figure*} 

Based on this convergence towards a stable area of the parameter space and the disappearance of the more chaotic population from the posterior distribution, a convenient choice for the NAFF stability threshold is the NAFF value above which most of the solutions would disappear after the addition of more observational constraints. In other words, looking at Fig. \ref{fig:DistribNAFF_HD45364} we choose this threshold as the bottom of the weakly chaotic population, i.e. the bottom of the blue distribution. In the case of HD\,45364, the limit value NAFF = -8.0 seems a convenient choice. 
There is of course some freedom in this decision. The important aspect to keep in mind though is to avoid rejecting configurations that are actually dynamically stable and thus constitute plausible solutions for the system's model. This should motivate us to select slightly less constraining NAFF limits. 
It is clear that the stability information will not impact the HPS+ posterior distribution, which is already highly constrained by the large amount of observations. 
In Table \ref{tab:Solution_37124_82943}, we update the planetary parameters from the analysis of the HPS+ that contains both CORALIE and HARPS measurements.

\subsection{Validation of the calibration strategy on HD\,202696}
To sum up, we propose the following NAFF calibration process for any given multi-planet system. 
\begin{enumerate}
\item Integrate a set of system's configurations derived e.g. from an MCMC exploration algorithm. Compute the NAFF indicator for each configuration that survived the entire integration. 
\item Based on the NAFF distribution, identify the value $x$ that
marks the start of the weakly chaotic distribution. Set up the stability threshold: NAFF$_{thr}$ = $x$. Any configuration with a NAFF smaller than this threshold is classified as stable. 
\end{enumerate}

It is important to note that the calibration strategy we introduced and illustrated here follows an exploratory approach. 
As a test of that calibration procedure and the stability-driven approach proposed in this work, we applied it to the two-planet system HD\,202696 for which long 2Myr integrations were performed. The NAFF chaos indicator in contrast was computed on the first 31\,250 yr (cf. Fig. \ref{fig:HD202696_NAFF-ST}). 
We present the histogram of the NAFF values in Appendix \ref{Appendix:NAFFdistributions}. It is composed of 4\,688 points, which are as many system's configurations that survived for at least 31\,250 years. That NAFF distribution harbours two closely-spaced peaks at stronger chaos separated by a third peak in the weakly chaotic regime. As the middle peak could potentially be the expression of stable bounded chaos, we decide to not rule it out. As such, we set the NAFF stability threshold at the bottom of that peak, and choose NAFF$_{thr}$=-4. A total of 2\,636 system's configurations make it to the NAFF-stable sample.

In order to assess the performance of this fast stability estimation, we compared its results with long integrations. Fig. \ref{fig:HD202696_DistribCompa} shows the parameters' distributions obtained in three different ways. In red, the original posteriors obtained from the set of RV measurements only, and composed of 7\,813 solutions. In black, the sub-sets of truly stable configurations, which survived the entire 2Myr integrations (2\,630 solutions). And finally in blue, the distributions of NAFF-stable configurations as derived after the first 31\,250 years of integration (2\,636 solutions). Both the dynamically-motivated black and blue distributions diverge from the original red distributions in the same way. Their high level of similarity confirms the equivalence of the stability-driven technique proposed in this work (blue distributions) with the "brute force" long integrations (black distributions). Furthermore, the gain in computing time is significant, with nearly two orders of magnitude in the case of HD\,202696. 
As a result, the parameters of the outermost planet are significantly refined with the stability-driven technique. Notably, its orbital eccentricity is damped. Consequently, we observe simultaneously a flattening in the distribution of its argument of periastron $\omega_c$, given the poorer definition of this parameter at smaller eccentricities. The parameters obtained from the NAFF-stable distributions are reported in Table \ref{tab:Solution_37124_82943}. 

In order to rigorously confirm this calibration technique, more highly well-constrained multi-planet systems of diverse architectures are needed, for which only the population of weakly chaotic configurations would remain in the posterior distribution after running a MCMC. This group of systems is nowadays elusive, but should grow little by little as more data are being gathered and new precise instruments are being built.

\section{Application to a system of Jovian planets: HD\,37124} 
\label{Sect:HD37124} 
HD\,37124 is currently known as a three-planet system. The planetary companions to this G4-type star (M=0.83M$_{\odot}$) were successively published in \citet{Vogt2000}, \citet{Butler2003} and \citet{Vogt2005}. In the latter, the authors used a set of 52 HIRES RV measurements to state the existence of three planets in the system. The model was still left unclear though, with a preference for three planets with orbital periods of 154.5, 844 and 2295 days, and all of them in the Jovian mass regime. Dynamical tests on this model showed significant planet-planet gravitational interactions, and a threshold of stability was set depending on the eccentricity of the outermost planet. However, the simulations did not exhaustively explore the parameter space, but a sub-space of it (period and eccentricity). In this sub-section, we aim at providing a revision of the orbital parameters on this system using the stability-driven approach introduced in Sect. \ref{Sect:Method}.  

We performed an iterative signal search in the periodogram on the same dataset than \citet{Vogt2005}, using the DACE platform. We found three significant periods close to the ones indicated here above, and we modelled each of them with a keplerian. We explored the parameters of this model with the MCMC described in \citet{Diaz2014} and implemented on DACE. Again, a burning phase of 25$\%$ was applied, and the posterior was further sampled according to the correlation length of the chain. 
We performed N-body simulations with \texttt{IAS15} over 50 kyr on a sample of 10\,000 solutions from the posterior. 
Out of these 10\,000 simulations, 4\,324 reached the end. 
More than half of the systems encountered an escape or close-encounter between two bodies. For each of the 4\,324 configurations, we computed the NAFF chaos indicator based on 20\,000 evenly spaced mean longitude values for each planet. The NAFF distribution of these configurations is presented in Appendix \ref{Appendix:NAFFdistributions}. Following the calibration procedure explained in Sect. \ref{Sect:NAFFcalib}, we set a stability threshold at NAFF = -2.5. As a result, the number of NAFF-stable configurations in our sample amounts to 2\,204. 

\begin{figure}
    \centering
\includegraphics[width=\columnwidth]{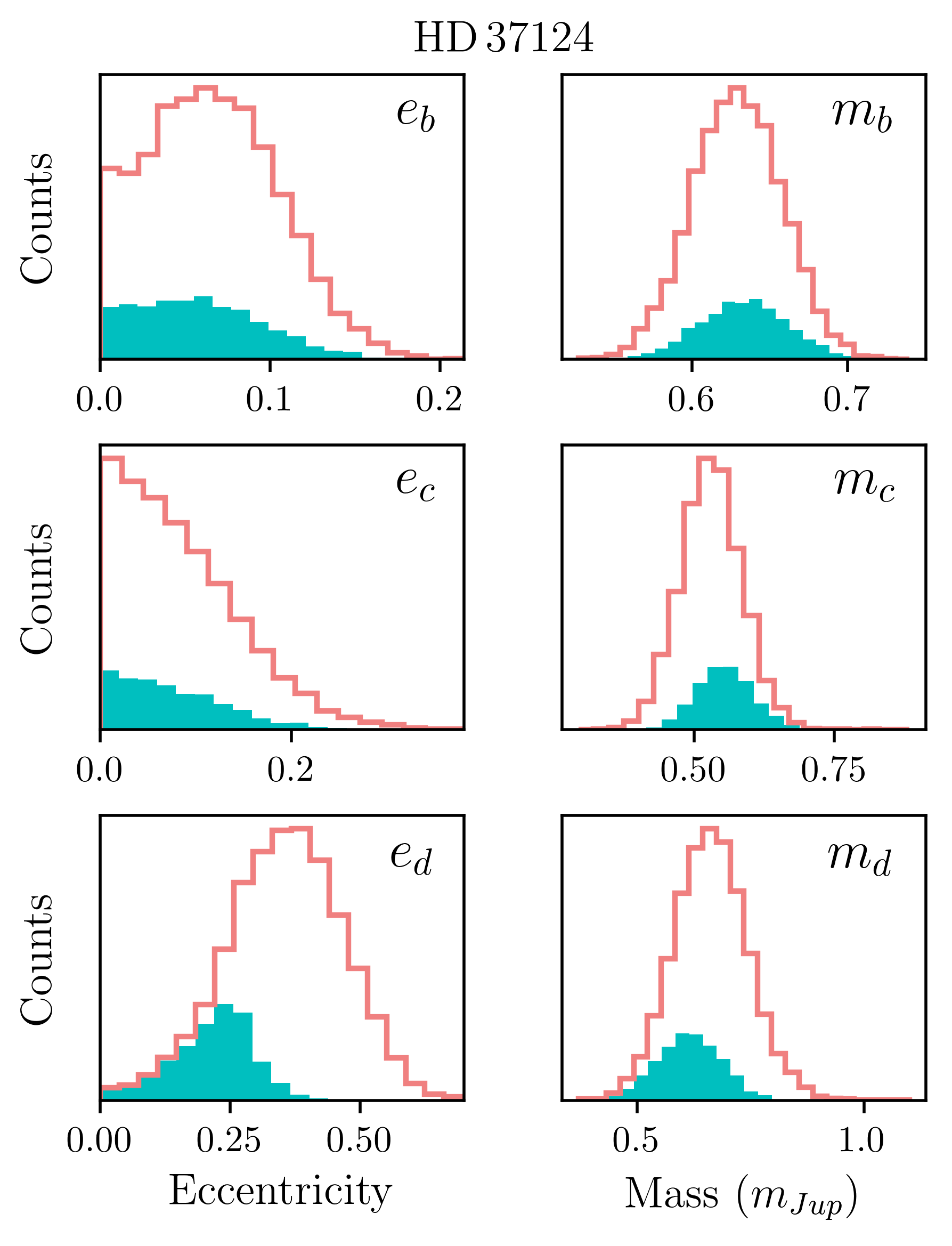}
\includegraphics[scale=1.0]{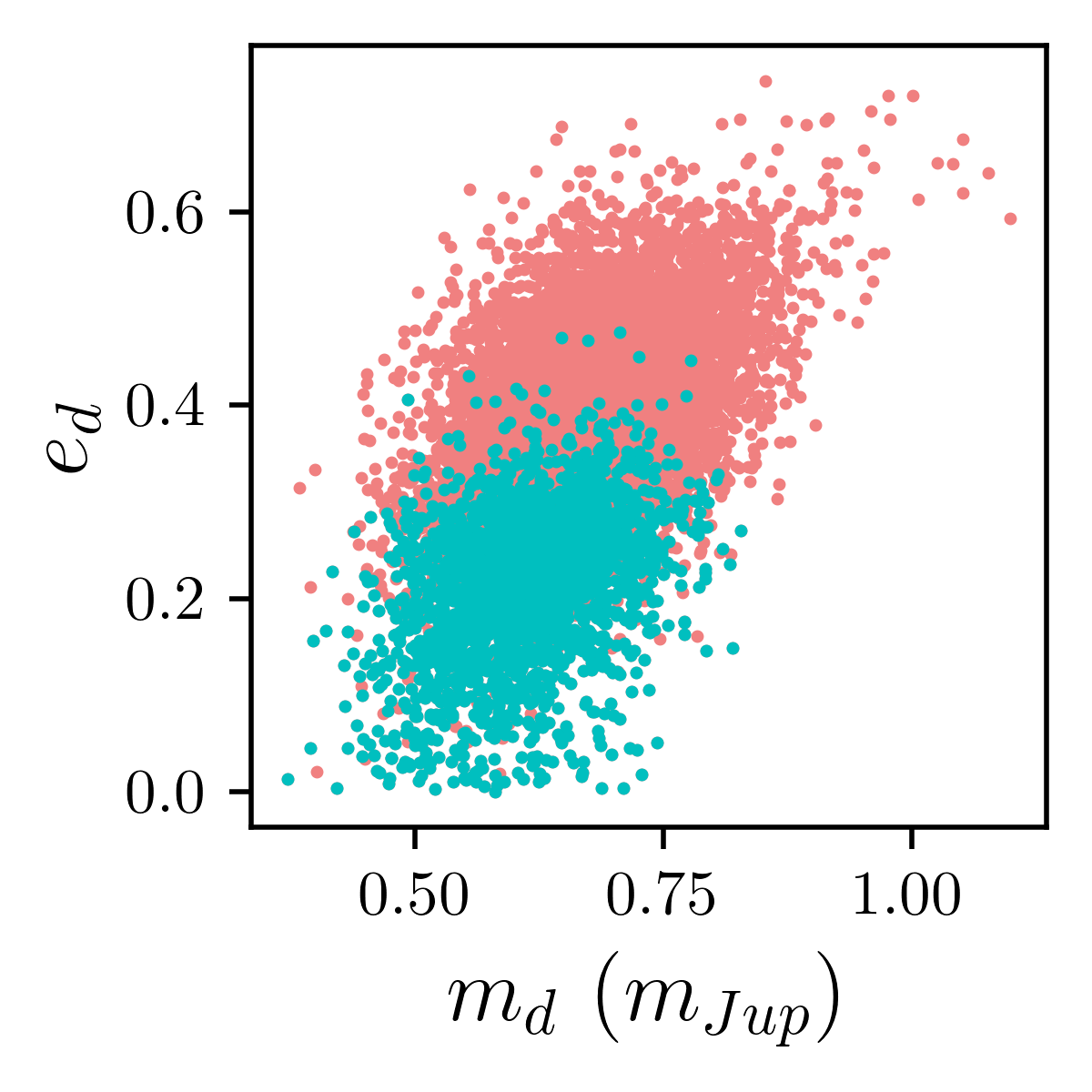}
    \caption{HD\,37124: Distributions of the orbital eccentricities and planetary masses both before (in solid red) and after (in blue) the inclusion of the orbital stability information.}
    \label{fig:HD37124}
\end{figure}

We  compare the distributions before and after the NAFF sorting. The main effects of the orbital stability constraint on HD\,37124 are observed on the orbital eccentricities and planetary masses. These results are presented in Fig. \ref{fig:HD37124}. In this grid of graphs, the left column lists the distributions of the eccentricity for each of the three planets, both for the full distributions (solid red lines - 10\,000 system configurations) and the distributions of NAFF-stable configurations only (in blue - sub-sample of 2\,204 configurations). For planets b and c, we observe that the stability-driven approach slightly favours the low eccentricities. This result is generally expected. The same behaviour appears for the outermost planet, but is strongly stressed. 
Indeed, the large eccentricities allowed by the MCMC posterior bring the outermost planet sufficiently close to the middle one such that planet-planet interactions are strong enough to destroy the system. The stability-driven approach is thus particularly sensitive to the orbital eccentricities. The right column presents the distributions for the planetary masses. While the mass estimation of the innermost planet is not significantly modified by the dynamical constraints, the same is not true for the two outer planets. 
Our stability-driven approach slightly increases the mass estimate of planet c, and at the opposite slightly decreases the mass estimate for planet d. The shift in the latter is the consequence of the correlation between $e_d$ and $m_d$ as shaped by the observational data. The bottom plot of Fig. \ref{fig:HD37124} shows the distributions projected on this 2D space. It stresses a positive correlation between $e_d$ and $m_d$. As the large eccentricity values are discarded, naturally this disfavours the large mass domain as well. In Table \ref{tab:Solution_37124_82943}, we provide the revised planetary parameters.

\section{Application to a compact system of low-mass planets: HD\,215152} 
\label{Sect:HD215152} 
HD\,215152 is a K3-type main-sequence star (M = 0.77 M$_{\odot}$) harbouring low levels of activity. 
A compact suite of four planets was discovered orbiting this star with the HARPS spectrograph \citep{Delisle2018}. With orbital periods of 5.76, 7.28, 10.86 and 25.2 days, the planets have estimated minimum masses of 2.0, 1.5, 2.7 and 3.5 M$_{\oplus}$, respectively. Such compact systems of low-mass planets are particularly challenging to find with RV measurements alone, and necessitate an extensive effort both in observations and data analysis. Nevertheless, given the small amplitudes of the RV variations that the planets apply on the star, some planetary parameters remain mostly unconstrained, notably the orbital eccentricities.

\begin{figure*}
    \centering
\includegraphics[width=\textwidth]{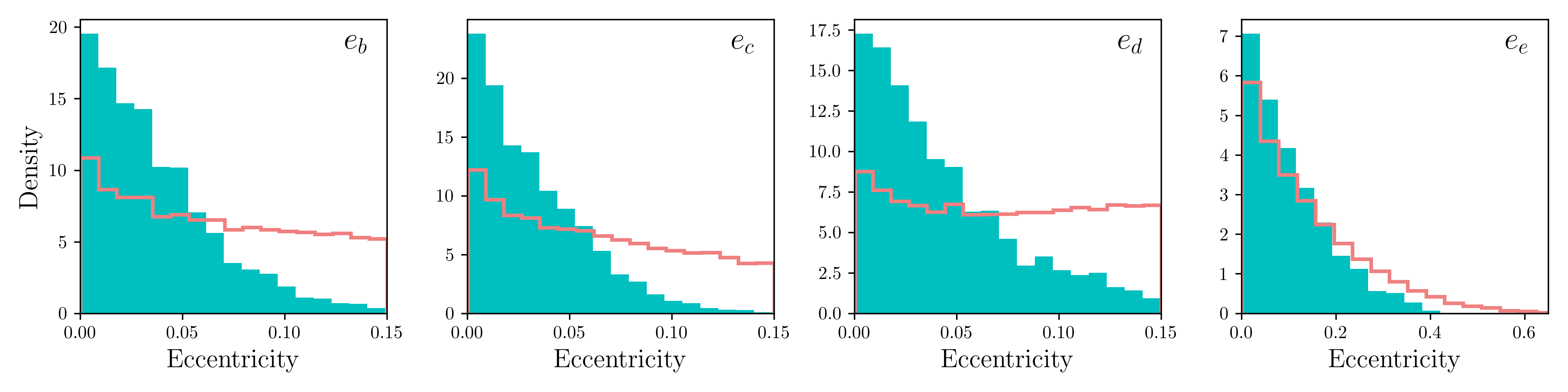}
    \caption{HD\,215152: posterior distribution projected on the orbital eccentricities (from left to right: innermost to outermost planet). The red line histograms represent the original posteriors, and are composed of 26\,250 system's configurations. The blue histograms are built up of the NAFF-stable solutions only, and consist in 3\,488 configurations.}
    \label{fig:Eccentricities_215152}
\end{figure*}

\citet{Delisle2018} explored the dynamics of the system with the computation of chaos maps. They showed that the two first planet pairs, involving the three innermost planets, lie outside (but close to) low-order mean motion resonances. Furthermore, these maps allowed the authors to place crude limits on the orbital eccentricity of the second planet $e_c$ for different overall system's inclination. A full exploration of the parameter space is necessary to provide rigorous dynamical constraints on the planetary parameters. 
We undertake this task in this section. However, the orbital inclination and longitude of ascending node are not constrained by the observations. Hence we limit our study to the co-planar edge-on configuration. 

We collected the posterior distribution of the eccentric model from \citet{Delisle2018}, composed of 7\,501 solutions. We computed the orbital evolution of each system's configuration over 5 kyr with the integrator \texttt{IAS15}. Again, on the simulations that reached the end (no close-encounter, no escape), we computed the NAFF fast chaos indicator based on a timeseries of 20\,000 equally spaced mean longitude measurements of each planet. Following the calibration procedure presented in Sect. \ref{Sect:NAFFcalib}, we set the stability threshold to NAFF\,=\,-1. At the end of this process, 140 solutions are classified as NAFF-stable and constitute the stable posterior distribution. This is too few to provide a reliable revision of the planetary parameters. This extreme proportion of unstable solutions is caused by the large exploration in the orbital eccentricities, mostly unconstrained by the RV measurements. 

As such, we performed a new exploration of the four-planet model with tighter constraints on the orbital eccentricities. To proceed, we gathered the HARPS data published in \citet{Delisle2018} and applied a similar correlated noise model SPLEAF with a Matérn kernel function \citep{Delisle2020}. We recovered the four known planets through an iterative search in a periodogram. We modelled them with keplerians, and explored the model parameters with a MCMC. The selected priors are mostly uninformative, except with the orbital eccentricities for which we imposed a Beta distribution with parameters $a$=0.867 and $b$=3.03, according to \citet{Kipping2013}. Furthermore, with the three innermost planets b, c and d that evolve in a tightly packed configuration, we cut those priors at 0.15, so to constrain the eccentricities in [0,0.15]. This choice is in accordance with the chaos maps presented in \citet{Delisle2018} (Fig. 6 of the cited paper), that show large levels of chaos above $e_c\sim$0.05 in the plausible domain of orbital periods. The MCMC we employed is an adaptive Metropolis algorithm described in \citet{Delisle2018}, so to ensure a derivation of the new posterior distribution equivalent to the one obtained by these authors. We performed 7M iterations and discarded the first 25$\%$ as burning phase. Eventually, 26\,250 solutions compose the posterior distribution. We computed the orbital evolution of each of these solutions over 5k years using the integrator \texttt{IAS15}. Out of the 26\,250 system configurations, 3\,938 reached the end of the simulation. The NAFF chaos indicator was estimated on each of them based on evenly sampled timeseries of the planetary mean longitudes. The NAFF distribution is presented in Annex \ref{Appendix:NAFFdistributions}. Two populations of systems, strongly and weakly chaotic, appear clearly. Following the NAFF calibration procedure, we set a stability threshold in-between these two groups, at NAFF=-2.0. As a result, 3\,488 system configurations are flagged as stable and constitute the NAFF-stable posterior distribution. Exactly 450 solutions were discarded with our NAFF criterion. The other removed solutions from the posterior were thus unstable during the N-body simulations, through either an encounter or escape of one of the planets. 

In this tightly packed system, the impact of the orbital eccentricities on the overall system's stability is significant. Despite the strong priors imposed on the distributions, the stability information further constrain the orbital eccentricities. Fig. \ref{fig:Eccentricities_215152} presents the posterior distribution projected on the orbital eccentricities of the planets, both before (solid red lines - 26\,250 solutions) and after (blue histograms - sub-group of 3\,488 configurations) the exclusion of NAFF-unstable systems. These density plots show a clear preference for the lowest values of the eccentricity. This trend is less strong on the outermost planet, due to its larger separation to the three innermost planets. 
Because of its compactness, the system HD\,215152 is located in a region of the parameter space densely populated with MMRs. Their overlap generates strong chaos, which causes the instability on short timescales of most of the system's configurations among our MCMC posterior.

\begin{table*}[t]
\centering
\begin{threeparttable}
\caption{HD\,45364, HD\,202696, HD\,37124 and HD\,215152: Planetary masses and orbital elements revised with the stability-driven approach. The values indicated in this table are the medians of the NAFF-stable distributions, accompanied with the 68.27$\%$ confidence intervals or the upper limits on them when the distributions are one-sided. 
In the case of HD\,45364, the new dataset of combined CORALIE + HARPS measurements generates a posterior of almost-exclusively stable solutions.}
\label{tab:Solution_37124_82943}
\footnotesize
\begin{tabular}{@{}llllll@{}}
\toprule
\textbf{Planet}  & P    & $\lambda$ & e & $\omega$ & m  \\ 
  & [days]    & [deg] &  & [deg] & [\mjup] \\ \midrule
\underline{\large{HD\,45364}} (M=0.82M$_{\odot}$)  \\ 
HD\,45364 b  & 227.583$\substack{+0.115 \\ -0.117}$ & 30.1$\pm$1.6 & 0.0209$\substack{+0.0196 \\ -0.0131}$ & 104.6$\substack{+54.8 \\ -73.1}$ & 0.184$\pm$0.008 \mjup \\ 
HD\,45364 c  & 344.135$\substack{+0.110 \\ -0.112}$ & 202.6$\pm$0.5 & 0.0290$\substack{+0.0085 \\ -0.0079}$ & 8.0$\substack{+20.8 \\ -26.9}$ & 0.574$\pm$0.024 \mjup \\ 
  &  &  &  &  &   \\ 
 \underline{\large{HD\,202696}} (M=1.91M$_{\odot}$)  \\ 
  HD\,202696 b  & 495.9$\substack{+19.9 \\ -16.3}$  & 179.4$\pm$4.5 & < 0.0774  & -27.7$\substack{+120.7 \\ -94.4}$ & 2.09$\pm$0.15 \mjup  \\ 
HD\,202696 c  & 969.8$\substack{+59.0 \\ -48.7}$ & 191.5$\substack{+6.7 \\ -7.5}$ & < 0.0866   & -11.0$\substack{+141.6 \\ -117.7}$ & 1.84$\pm$0.19 \mjup  \\ 
  &  &  &  &  &   \\ 
 \underline{\large{HD\,37124}} (M=0.83M$_{\odot}$)  \\ 
HD\,37124 b  & 149.590$\substack{+0.078 \\ -0.097}$ & 5.0$\substack{+5.2 \\ -3.5}$ & 0.0556$\substack{+0.0410 \\ -0.0370}$ & 123.1$\substack{+31.8 \\ -116.9}$ & 0.631$\pm$0.029 \mjup  \\ 
HD\,37124 c  & 842.8$\substack{+10.5 \\ -5.8}$ & -12.3$\substack{+9.3 \\ -17.1}$ & <0.137 ; <0.309(99.73$\%$) 
& 9.2$\substack{+98.2 \\ -81.1}$ & 0.549$\pm$0.051 \mjup  \\ 
HD\,37124 d  & 2123.6$\substack{+127.6 \\ -195.7}$ & 145.8$\substack{+58.4 \\ -34.2}$ & 0.2530$\substack{+0.0668 \\ -0.0920}$  & -74.4$\substack{+36.5 \\ -29.2}$ & 0.623$\pm$0.071 \mjup  \\ 
 &  &  &  &  &   \\ 
 \underline{\large{HD\,215152}} (M=0.77M$_{\odot}$)  \\ 
 HD\,215152 b  & 5.76010(84)  & 169.7$\pm$20.5  & <0.0660 & -7.1$\substack{+109.3 \\ -103.6}$  & 1.65$\pm$0.29 \mearth  \\ 
 HD\,215152 c  & 7.2826$\substack{+0.0025 \\ -0.0033}$ & 91.4$\substack{+31.2 \\ -36.1}$  & <0.0579 & -18.2$\substack{+126.8 \\ -103.4}$ & 1.57$\substack{+0.32 \\ -0.36}$ \mearth  \\ 
 HD\,215152 d  & 10.8661(28) & 98.6$\pm$18.4 & <0.0794 & 209.9$\substack{+87.7 \\ -108.3}$ &  2.54$\pm$0.43 \mearth \\ 
 HD\,215152 e  & 25.191(23) & 71.5$\pm$27.6 & <0.1880  & 172.5$\substack{+122.9 \\ -103.7}$ &  2.38$\pm$0.65 \mearth  \\ \bottomrule \bottomrule
\end{tabular}

\bigskip

\begin{tabular}{@{}rlrl@{}}
\textbf{Notation}  & \textbf{Description} & \textbf{Notation}  & \textbf{Description}  \\ \midrule
P & Orbital period  & $\omega$ & Argument of periastron \\ 
$\lambda$ & Mean longitude & m & Planetary mass  \\ 
$e$ & Orbital eccentricity & & \\ 
\end{tabular}

\end{threeparttable}
\end{table*}

\section{Conclusions} 
\label{Sect:Conclusions} 
\begin{figure}
    \centering
\includegraphics[width=\columnwidth]{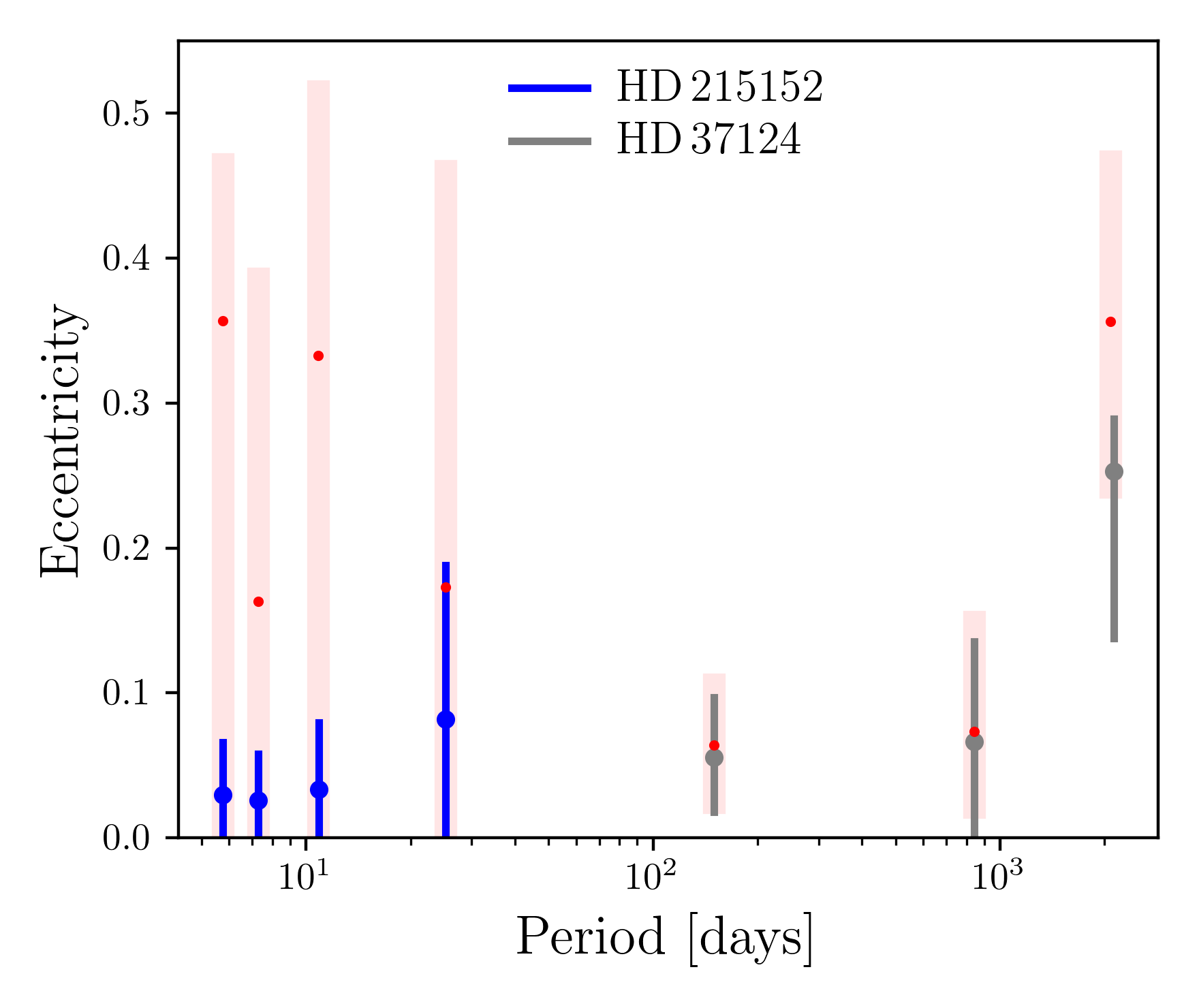}
    \caption{HD\,215152 and HD\,37124 plotted in the Period - Eccentricity sub-space together with the 68.23$\%$ confidence intervals on the eccentricities, both before (light red) and after (blue and grey) applying the stability-driven approach.}
    \label{fig:215152-37124_P-e}
\end{figure}

We developed a method that aims at refining the planetary masses and orbital parameters in multi-planet systems via a global inclusion of dynamical constraints. Such an approach makes use of the NAFF fast chaos indicator, for which we propose a general calibration strategy. The latter needs to be iterated for each significantly different architecture (i.e., for each new multi-planet system under study), but does not necessitate additional computational cost. We illustrated our calibration strategy on HD\,45364 and presented new RV measurements that further point towards the 3:2 MMR state for this system. Our calibration process was validated onto HD\,202696, by comparing the results of the approach with long 2Myr numerical integrations. We then applied this technique on HD\,37124, a system composed of three Jovian planets. We demonstrated the potential of the technique to refine the planetary parameters, and especially the orbital eccentricities and the planetary masses. Furthermore, given the global exploration of the parameter space that this technique uses, we could observe important correlations between parameters that explained the origin of some of the dynamical constraints. The stability-driven approach that we propose helps to understand the dynamical state of the system. Finally,  we applied the stability constraint on HD\,215152, a four-planet system in a very compact configuration. Again, we observe a significant influence of the orbital eccentricities on the overall system's stability, which allows to put strong and rigorous constraints on their upper values. 
Such an impact on the orbital eccentricities is synthesised in Fig. \ref{fig:215152-37124_P-e}. In that plot, the planets in HD\,37124 and HD\,215152 are presented in the (P,e) space together with their 68.23$\%$ confidence intervals on the eccentricity, both before (light red) and after (blue and grey) the update with the stability-driven refinement technique. The new views on these systems are in better agreement with the expected results from formation processes, which generally produce low eccentric orbits due to the damping effects in the proto-planetary disk. Those updates, when generalised to the exoplanet population level, can potentially bring a new light on the formation and evolution models.

Overall, the stability-driven approach that we presented in this work is faster than the "brute force" numerical integrations, and provides reliable constraints on the systems under study. Some caveats remain in the NAFF calibration procedure, and in particular the current lack of highly precise multi-planet systems. A growing number of them should help us to confirm the trend that we see in the systems' posterior distributions, regarding the proportion of strongly chaotic system's configurations with respect to the precision of observational data, and which is at the basis of our calibration strategy. Let us note that in this work, the impact of orbital inclination and in particular mutual inclination was not explored. The orbital inclinations of the planets in all the systems studied here were fixed at 90 degrees. The study of TOI-125 in \citet{Nielsen2020} explored the impact of mutual inclination on the overall system's stability, based on our technique. An exhaustive presentation of this impact would prepare our approach for the inclined systems. While the very small mutual inclinations between the transiting planets should not have a strong dynamical influence, this is not the case of the exoplanets discovered via astrometry. Notably, the GAIA mission is expected to provide the community with numerous massive exoplanet detections on wide orbits, together with the information about their orbital inclination. 
Concerning smaller non-transiting planets closer-in to their host star, the strong planet-planet interactions that some of them experience may unveil their orbital inclination too. Indeed, the variations of the orbital elements could be measured in the radial velocity timeseries using dynamical fits, lifting the degeneracy between the mass and the orbital inclination. This has been achieved in a few cases, and requires high signal-to-noise RVs and long baselines of observations \citep[e.g.][with HD\,82943]{Tan2013}. 

Finally, we emphasise the importance to carry a systematic dynamical revision of newly discovered multi-planet systems, and in particular for those in compact configurations, as the stability information can potentially significantly refine planetary parameters. This is of importance in light of selecting the best target candidates for future missions on the next-generation telescopes such as James Webb Space Telescope (JWST), PLATO or the E-ELT.

\begin{acknowledgements}  
We thank the anonymous referee for his/her valuable comments and suggestions, which helped improve the quality of this manuscript. 

This work has been carried out in the frame of the National Centre for Competence in Research PlanetS supported by the Swiss National Science Foundation (SNSF). This project has received funding from the European Research Council (ERC) under the European Union's Horizon 2020 research and innovation programme (project {\sc Spice Dune}, grant agreement No 947634). 

This publication makes use of the Data $\&$ Analysis Center for Exoplanets (DACE), a platform of the Swiss National Centre of Competence in Research (NCCR) PlanetS, federating the Swiss expertise in Exoplanet research. 

Tools used: \texttt{REBOUND} \citep{Rein2012} , \texttt{SPLEAF} \citep{Delisle2020}
\end{acknowledgements}

\bibliographystyle{aa} 
\bibliography{bib.bib}

\appendix
\section{Data analysis of HD\,45364} 
\label{Appendix:HD45364_DataAnalysis}
\begin{figure}
    \centering
\includegraphics[width=\columnwidth]{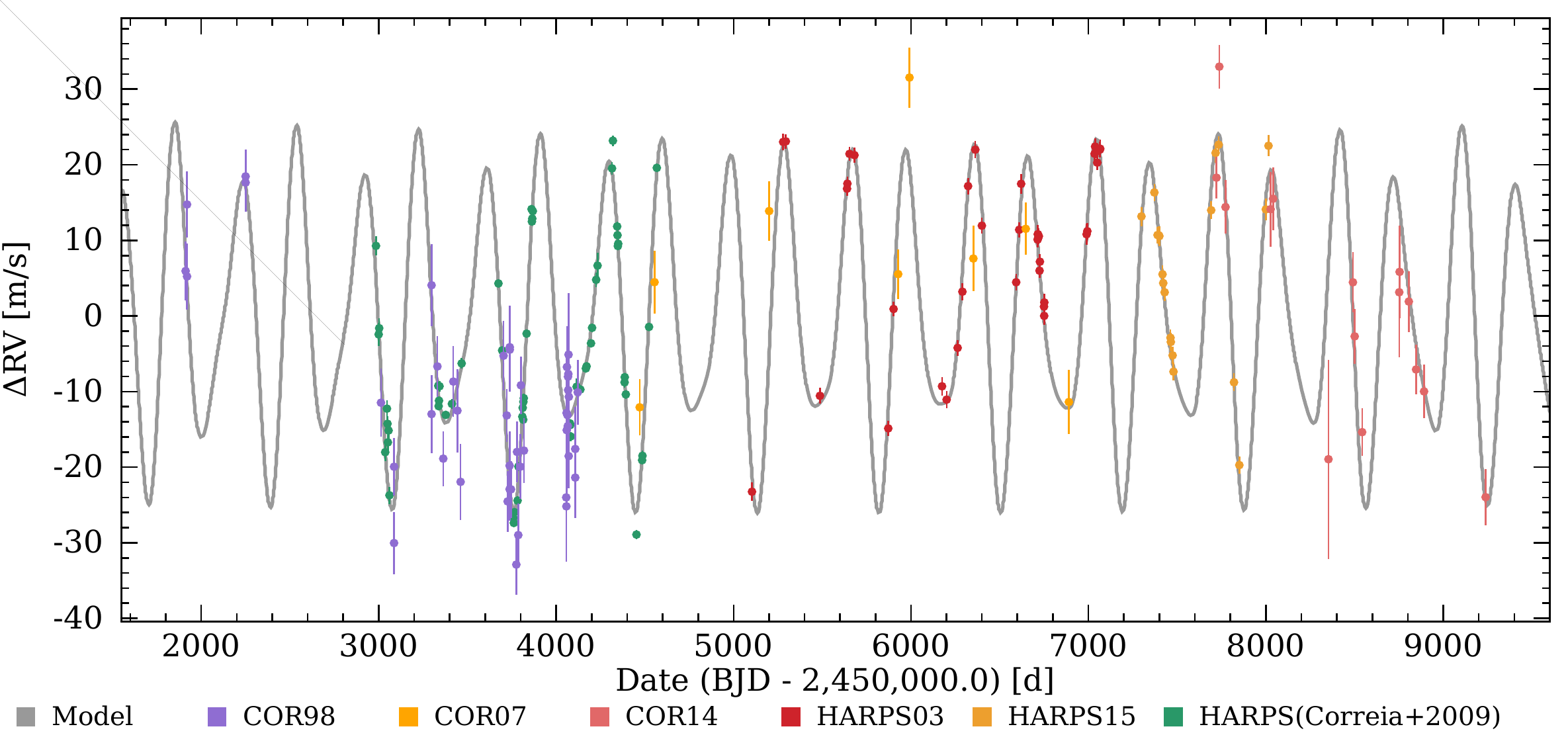}
\includegraphics[width=\columnwidth]{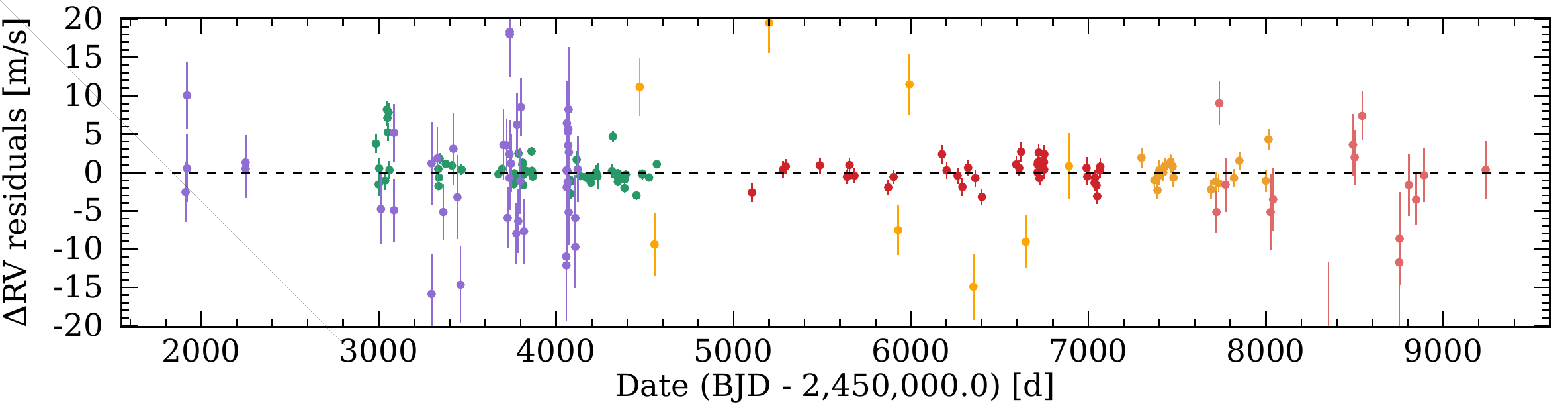}
    \caption{HD\,45364: Full RV timeseries. It contains the already-published HARPS data (in green), the additional HARPS data gathered after 2009 (HARPS03 and HARPS15) and the CORALIE data covering 20 years of observations (COR98, COR07 and COR14). Finally, the grey curve presents the best-fit of the two-keplerian model.}
    \label{fig:HD45364_RVtimeseries}
\end{figure}

\begin{figure}
    \centering
\includegraphics[width=\columnwidth]{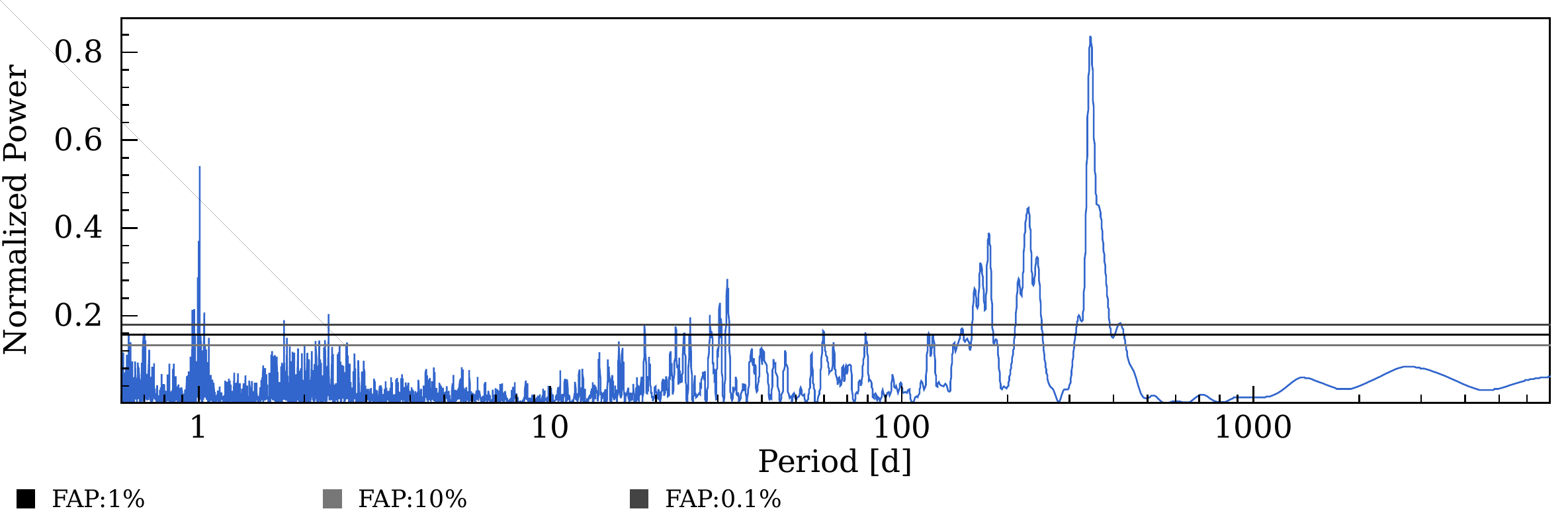} \\ 
\includegraphics[width=\columnwidth]{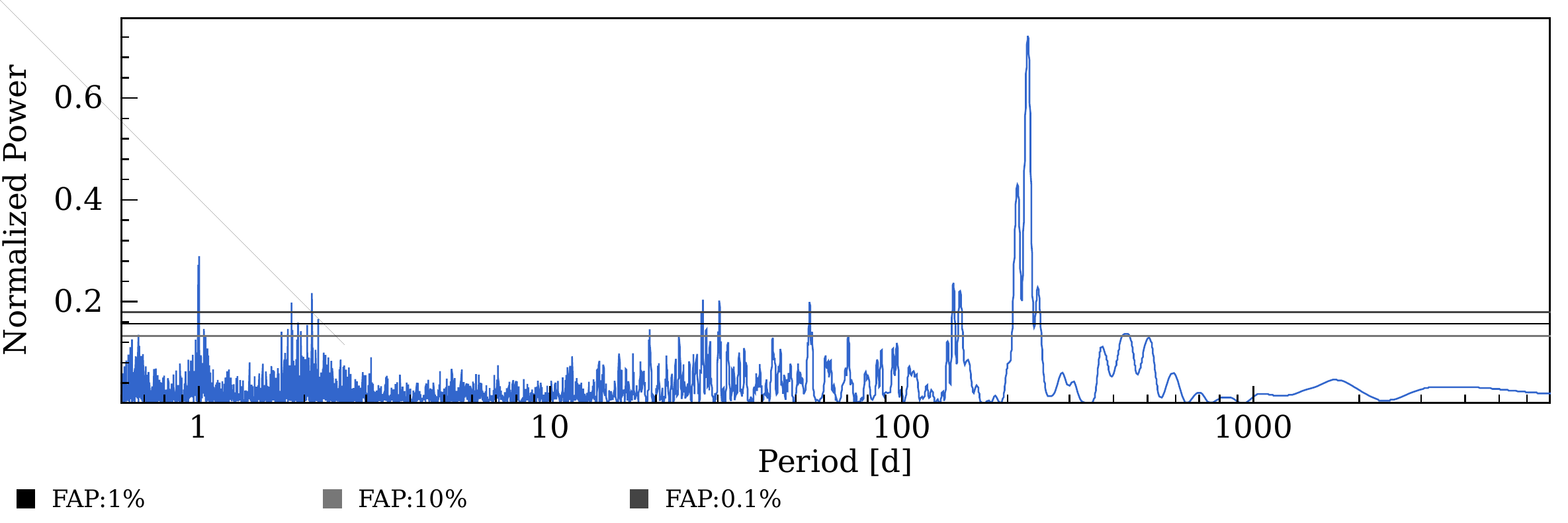} \\ 
\includegraphics[width=\columnwidth]{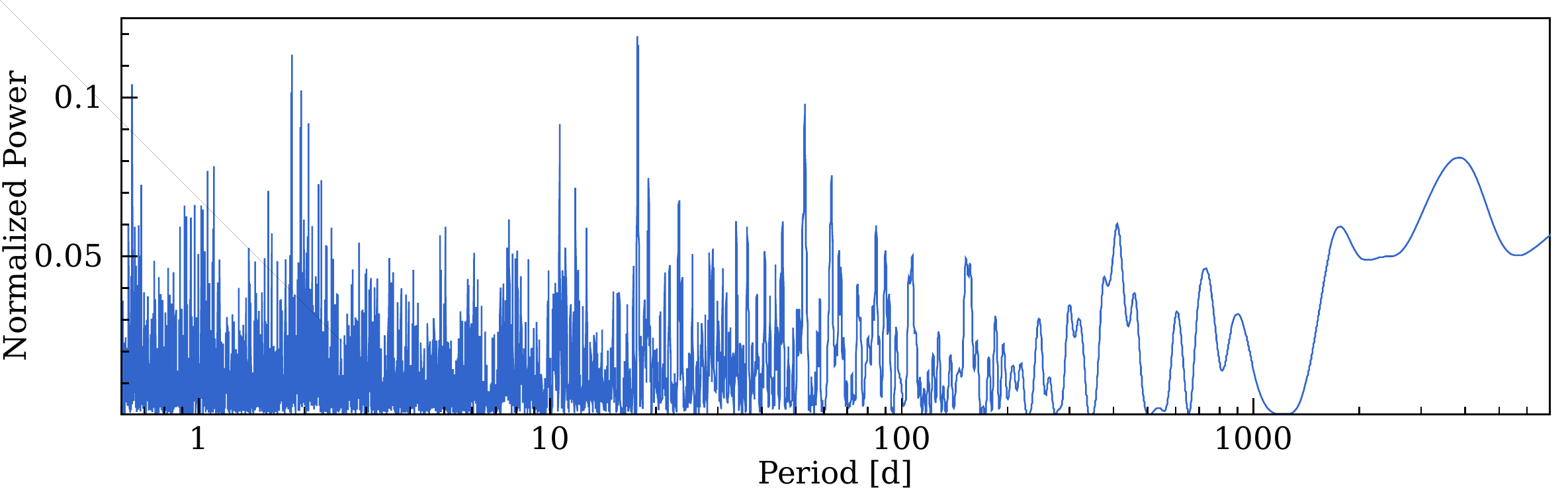} \\ 
    \caption{HD\,45364: Periodograms of the RV timeseries. The three horizontal lines correspond to different levels of False Alarm Probability (FAP), of 10$\%$, 1$\%$ and 0.1$\%$ from bottom to top. \textit{Top} : Periodogram of the RV timeseries with no keplerian in the model. The highest peak is very significant and corresponds to a signal of period 344.03 days. \textit{Middle} : Periodogram of the residuals after subtraction of that signal, modelled with a keplerian. Another highly significant signal is detected, at period = 227.69 days. \textit{Bottom} : Periodogram of the residuals after subtraction of the two-keplerian model.}
    \label{fig:HD45364_RVperiodograms}
\end{figure}

We present a new yet-unpublished set of RV measurements of the star HD\,45364. It consists in 57 measurements taken with HARPS between Sept 29, 2009 and Sept 18, 2017. Also included are 68 measurements taken with CORALIE between Jan 2, 2001 and Jan 24, 2021. These measurements are less precise than HARPS, but they have the advantage to increase the baseline of observations. These two sets add to the already published HARPS measurements \citep{Correia2009}. The new HARPS and CORALIE data are further separated according to significant upgrades in the instruments, giving rise respectively to HARPS03 and HARPS15, and COR98, COR07 and COR14. The COR98 radial velocities harbour a strong correlation with the bissector span of the Cross-Correlation Function (CCF). We corrected for it through a detrending with a Gaussian kernel. Then we performed a periodic signal search in the Lomb-Scargle periodogram of the full dataset, via the platform DACE, and fitted these periodicities with keplerians. Two keplerians corresponding to the known planets were included in the model, after which no significant signal remains in the residuals. Hence we see no hint of additional planet in the system. In Fig. \ref{fig:HD45364_RVtimeseries}, we present the updated RV timeseries together with the two-keplerian model's best fit. Fig. \ref{fig:HD45364_RVperiodograms} traces back the elaboration of the two-keplerian model, via a successive signal search in the Lomb-Scargle periodogram. 

\section{NAFF distributions for HD\,202696, HD\,37124 and HD\,215152} 
\label{Appendix:NAFFdistributions}
\begin{figure*}
    \centering
\includegraphics[width=\textwidth]{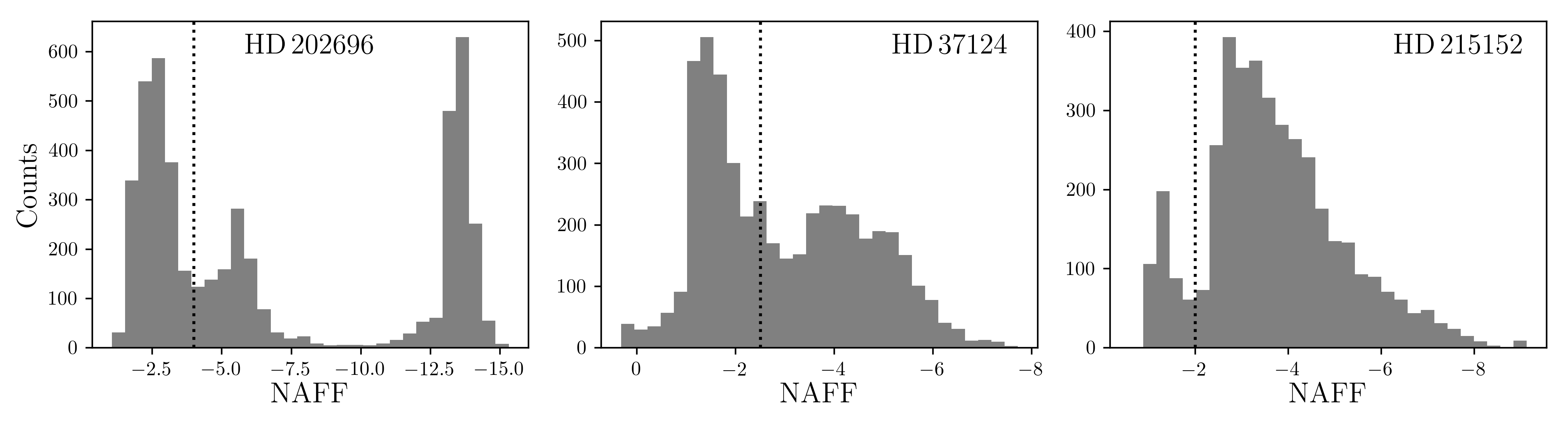}
    \caption{NAFF distributions for the posteriors of HD\,202696, HD\,37124 and HD\,215152 (left, middle and right). These distributions are composed of 4\,688, 4\,324 and 3\,938 system configurations, respectively.}
    \label{fig:NAFFdistribs_202696-37124-215152}
\end{figure*}

We present in this Appendix the posterior distributions of HD\,202696, HD\,37124 and HD\,45364 projected onto the NAFF chaos indicator. They were obtained from an exploration of the parameter space of each system's model, and dynamical simulations on each solution of the samples. Every single point composing the NAFF distributions hence corresponds to a solution for the system that survived as well to the entire short-term N-body simulation in order to estimate its NAFF chaos level. The NAFF distributions of the systems HD\,202696, HD\,37124 and HD\,215152 are displayed in Fig. \ref{fig:NAFFdistribs_202696-37124-215152}. 

These distributions serve to assign the NAFF-stability threshold of each system. Since they are the by-product of the stability-driven approach, the NAFF calibration does not necessitate additional computational cost. The threshold separates the weakly from the strongly chaotic populations of solutions (cf. Sect. \ref{Sect:NAFFcalib}). Hence, from the histograms above, we set stability thresholds to the following: for HD\,202696: NAFF=-4.0, for HD\,37124: NAFF=-2.5, for HD\,215152: NAFF=-2.0. They are spotted in Fig. \ref{fig:NAFFdistribs_202696-37124-215152} with the vertical dotted lines.

\end{document}